\documentclass[notitlepage,letterpaper,preprint]{article}

\usepackage{amsmath}
\usepackage{amsfonts}
\usepackage{amssymb}
\usepackage{graphicx}
\usepackage{subfig}
\usepackage{cancel}
\usepackage{booktabs}
\usepackage{color}
\usepackage{cite}
\usepackage[colorlinks=true,urlcolor=blue,linkcolor=blue,citecolor=blue]{hyperref}
\usepackage[top=1cm, bottom=2cm,outer=2cm, inner=2cm]{geometry}

\begin{document}

\title{Convection and cracking stability of spheres in General Relativity}
\author{
\textbf{H\'ector Hern\'andez}\thanks{\texttt{hector@ula.ve}} \\
\textit{ Escuela de F\'isica, Universidad Industrial de Santander,}\\ 
\textit{Bucaramanga 680002, Colombia} and \\
\textit{Grupo de F\'{\i}sica Te\'orica, Departamento de F\'{\i}sica,} \\ 
\textit{Universidad de Los Andes, M\'{e}rida 5101, Venezuela.} \\
\textbf{Luis A. N\'{u}\~{n}ez}\thanks{\texttt{lnunez@uis.edu.co}} \\
\textit{ Escuela de F\'isica, Universidad Industrial de Santander,  }\\ 
\textit{Bucaramanga 680002, Colombia} and \\
\textit{Centro de F\'{\i}sica Fundamental, Departamento de F\'{\i}sica,} \\ 
\textit{Universidad de Los Andes, M\'{e}rida 5101, Venezuela.}; \\
 and 
 \textbf{Adriana V\'asquez-Ram\'irez}\thanks{\texttt{adrianacvr67@gmail.com}} \\
\textit{ Escuela de F\'isica, Universidad Industrial de Santander,  }\\ 
\textit{Bucaramanga 680002, Colombia} 
}
\maketitle

\begin{abstract}
In the present paper we consider convection and cracking instabilities as well as their interplay. We develop a simple criterion to identify equations of state unstable to convection, and explore the influence of buoyancy on cracking (or overturning) for isotropic and anisotropic relativistic spheres. We show that a density profile $\rho(r)$, monotonous, decreasing and concave , i.e. $\rho' < 0$ and $\rho'' < 0$, will be stable against convection, if the radial sound velocity monotonically decreases outward. We also studied the cracking instability scenarios and found that isotropic models can be unstable, when the reaction of the pressure gradient is neglected, i.e. $\delta \mathcal{R}_p = 0$; but if it is considered, the instabilities may vanish and this result is valid, for both isotropic and anisotropic matter distributions.

\end{abstract}

\section{Introduction}
The stability of general relativistic self-gravitating matter distributions has been extensively studied and reported in the literature through several techniques for many years. It is complex multivariate  problem which depends on the micro-physics --bulk/shear viscosity, crust on the surface, magnetic field and so on-- of the material constituents and their description through a macroscopic equation of state that characterizes the configuration (see standard texts in Relativistic Astrophysics and Neutron Stars \cite{ShapiroTeukolsky1983,Demianski1985,Glendenning2000,HaenselPotekhinYakovlev2007,KippenhahnWeigertWeiss2013} and references therein). 

The studies of stability distinguish two different approaches associated to the global and local scales, where instabilities affect the structure. On the global scale it is examined through the dynamical perturbation scheme which, in the case of spherical symmetry, can be translated into how radial pulsations induce possible disruptions of a stellar body. On the other hand, local stability investigates the effects of convection and/or cracking of the material within the matter distribution. Pulsation is a global phenomenon characterized by the collective motion of the entire body while, convection/cracking occurs locally and is governed by the nearby values of the thermodynamical variables and their gradients (see an interesting review in \cite{FriedmanStergioulas2014}).

The dynamical instability approach studies the evolution of perturbations on the physical and geometrical variables throughout the matter distribution. In General Relativity, it arose from the seminal works of S. Chandrasekhar, R.F. Tooper and J.M. Bardeen \cite{Chandrasekhar1964a,Chandrasekhar1964b, Tooper1964b,Tooper1965,Bardeen1965} and, a decade later, was formalized  by  J. L. Friedman and B. F. Schutz \cite{FriedmanSchutz1975}. For an anisotropic fluid, this criterion bounds the adiabatic index as 

\begin{equation}
\label{GammaStability}
\Gamma = \frac{\rho + P}{P} v^{2} \geq \frac{4}{3} \, ,
\end{equation}
where $\rho$ denotes the energy density, $P$ the radial pressure, and $v^{2}$ the radial sound speed, respectively \cite{HeintzmannHillebrandt1975,ChanHerreraSantos1993,HerreraSantos1997,Ivanov2017}. Within the global stability criteria we can also identify the Harrison-Zeldovich-Novikov condition, which implies that $\mathrm{d}M(\rho_c) /\mathrm{d} \rho_c \geq 0$, where $M$ is the total mass of the configuration and $\rho_c$ the central density, of the distribution \cite{Ivanov2017}.

The stability of a spherical star to convection implies the buoyancy principle which leads to that pressure and energy density must, decrease outwards in any hydrostatic matter configuration \cite{Bondi1964B,Thorne1966,Kovetz1967}. Finally, the cracking instability approach determines the tidal acceleration profiles generated by perturbations of the  energy density and the anisotropy of pressures identifying the changes sign of the total force distribution within the system \cite{Herrera1992,DiPriscoEtal1994,DiPriscoHerreraVarela1997,AbreuHernandezNunez2007a}. This approach has been applied to an anisotropic fluid with barotropic equations of state \cite{AbreuHernandezNunez2007b} and, more recently extended to take into account the perturbation of the pressure gradient in both isotropic and anisotropic matter configurations \cite{GonzalezNavarroNunez2015,GonzalezNavarroNunez2017}. 

In the present paper we consider convection and cracking instabilities as well as their interplay. We develop a simple criterion to identify equations of state unstable to convection, and also explore the influence of buoyancy on cracking (or overturning) of isotropic and anisotropic relativistic spheres.   

This paper is organized as follows: Section \ref{EcuacionesGenerales} describes the general equations of the theoretical framework of General Relativity. 
In Section \ref{convection}  we formulate the concepts of adiabatic stability while cracking for self-gravitating anisotropic matter configurations is discussed in Section \ref{EffectsAnisotropic}. Next we present in Section \ref{AceptabilityConditions} the acceptability conditions which make any model physically reasonable. The models used and a discussion of our results for isotropic and anisotropic cases are presented in the Sections \ref{Models} and \ref{Modeling}. Finally, in Section \ref{FinalRemarks} we wrap-up our final remarks.

\section{The field equations}
\label{EcuacionesGenerales}
Let us consider a spherically symmetric  space-time whose line element is given by
\begin{equation}
\mathrm{d}s^2 = -e^{2\nu(r)}\,\mathrm{d}t^2+e^{2\lambda(r)}\,\mathrm{d}r^2+
r^2 \left(\mathrm{d}\theta^2+\sin^2(\theta)\mathrm{d}\phi^2\right),
\label{metricSpherical}
\end{equation}
with the regularity conditions at 
$r=0: e^{2\nu(0)}= \mbox{const.}, \,\, e^{-2\lambda(0)}= 1, \,\, {\nu}'(0)=\lambda'(0)=0\,.
$

At the surface of the sphere $r=r_b$, the interior solution should match continuously the exterior Schwarzschild solution, which implies that: $e^{2\nu(r_b)}=e^{-2\lambda(r_b)}=1-2\mu_b$, and the compactness at the surface defined as $\mu_b=2M/r_b$.

We shall consider a distribution of matter consisting of a non-Pascalian fluid with the energy-momentum tensor:
\begin{equation}
T_\mu^\nu = \mbox{diag}\left[\rho(r),-P(r),-P_\perp(r),-P_\perp(r)  \right] \,,
\label{tmunu}
\end{equation}
with energy density  $\rho(r)$, radial pressure $P(r)$ and tangential pressure $P_\perp(r)$ of the fluid being determined by the Einstein field equations as
\begin{eqnarray}
\rho(r)&=& \frac{ e^{-2\lambda}\left(2 r \lambda^{\prime}-1\right)+1 }
{8\pi r^{2}}\,,\label{ee1} \\
P(r) &=&  \rho(r) - \frac{e^{-2\lambda}}{4\pi r^{2}}\left[ r\left( \lambda^{\prime}-\nu^{\prime}\right)+ e^{2\lambda}-1\right]\,\label{ee2} \qquad \textrm{and} \\
P_\perp(r) &=&-\frac{e^{-2\lambda}}{8\pi}\left[ \frac{
\lambda^{\prime}-\nu^{\prime}}r-\nu^{\prime \prime }+\nu^{\prime}\lambda^{\prime
}-\left(\nu^{\prime}\right)^2\right] \,, 
\end{eqnarray}
where primes denote differentiation with respect to $r$.

As it is well known, $T^{\mu}_{r \; ; \mu} =0$ implies the hydrostatic equilibrium equation, or Tolman-Oppenheimer-Volkoff (TOV) equation  which, if $m(r)=\frac{r}{2} \left(1- e^{-2\lambda} \right)$, can be written for this anisotropic fluid as  
\begin{equation}
\frac{\mathrm{d} P}{\mathrm{d} r} +(\rho +P)\frac{m + 4 \pi r^{3}P}{r(r-2m)} -\frac{2}{r}\left(P_\perp-P \right) =0\,,
 \label{TOVStructure1}
\end{equation}
and together with
\begin{equation}
\label{MassStructure2}
\frac{\mathrm{d} m}{\mathrm{d} r}=4\pi r^2 \rho \,,
\end{equation}
constitute the stellar structure equations. In order to obtain the density and pressure profiles we have to provide two equations of state, $P=P(\rho)$ and $P_{\perp}=P_{\perp}(\rho)$, which for the present work will be assumed barotropic.

\section{Adiabatic convection stability condition}
\label{convection}
The stability of a spherical star against convection can be easily  understood. When a fluid element is displaced downward, if its density, $\rho_{e}$, increases more rapidly than the surrounding  density, $\rho_{s} $, the element will sink downward and the star will be unstable. On the other hand, if the density of fluid element is less than its surroundings, it will float back and the star will be stable to convection.

Thus we can single out these three cases: 
\begin{enumerate}
\item If $ \rho_{e}> \rho_{s} $, gravity will tend to push the fluid element downward further and the system will be unstable.
\item If $ \rho_{e} = \rho_{s} $, the system is considered neutral or metastable. 

\item If $ \rho_{e} <\rho_{s} $, a restoring force will act on the fluid element and then the system will be stable because it tends to its original state.
\end{enumerate}

Following Bondi \cite{Bondi1964B}, let us denote the density, $\rho (r_p)$, of an infinitesimal fluid element at its original position $r_p$  and displace this piece of material downward, thus:
\begin{equation}
\label{rhodesplazada}
\rho(r_p) \rightarrow \rho(r_p)+\delta \rho(r)\,,\,\, \textrm{with} \,\,\,
\delta \rho(r)=\rho'(r)(-\delta r) \,\, \textrm{and} \,\,\, r=r_p-\delta r\,,
\end{equation} 
where $r$ represents the current position of the fluid element, $r_p $ its original position and $-\delta r$ the downward shift. 

Because $\rho'(r) <0$, then $ \delta \rho(r) $ is a positive quantity, and the density of the compressed fluid element at the new displaced position will be greater that the density at its original position  $r_p$. 
On the other hand, expanding the density of the  environment at the displaced position we get:
\begin{equation}
\label{rhoentorno}
\rho(r_p-\delta r) \approx \rho(r_p)+\rho'(r_p)(-\delta r) \, ,
\end{equation}   

The system will be stable against convection if the environment density is greater or equal than the density of the fluid element, we then have:
\begin{equation}
\rho(r_p)+\rho'(r_p)(-\delta r) \geq \rho(r_p)+\rho'(r)(-\delta r),
\end{equation}
thus
$
\rho'(r_p)\leq \rho'(r) \, .
$

Now, expanding $\rho'(r)$ around $r_p$ we get
\begin{equation}
\label{BouyancyCriterion}
\rho'(r_p)+\rho''(r_p)\delta r \leq \rho'(r_p) 
\,\, \Rightarrow \,\, \rho''(r) \leq 0,
\end{equation}
which becomes the criterion of adiabatic stability against convection. Thus, density profiles with the second derivative less or equal than zero, $\rho''(r) \leq 0$, will be stable against adiabatic convective motions. 

It is clear that parabolic density profiles, $\rho = \alpha r^2 + \beta$, with $\alpha$ and $\beta$ constants, will be stable against this type of convection, because  buoyancy condition must also be fulfilled at the center $r = 0$  of the sphere: $\rho'< 0 \,\, \Rightarrow  \, \rho'_c=0,$ and $ \rho''_c<0 $. These profiles have been implemented for the MIT Bag model through a linear equation of state, $P = \beta(\rho -\rho_{s})$, when densities become high enough for a phase transition to quark matter to occur \cite{Glendenning2000,Ivanov2017}. 

\section{Convection and cracking sources of instability}
\label{EffectsAnisotropic}
In this section we shall consider convective instabilities in the framework of cracking induced by perturbation in the density profile. Just for completeness we outline here the main concepts and equations concerning cracking for isotropic and anisotropic matter configurations, for further details, we refer interested readers to  \cite{GonzalezNavarroNunez2015,GonzalezNavarroNunez2017} and references therein. 

As in the previous works \cite{GonzalezNavarroNunez2015,GonzalezNavarroNunez2017}, we assume that density fluctuations induce variations into all other physical variables, i.e.  $m(r), P(r), P_\perp(r)$ and their derivatives, generating a non-vanishing total radial force distribution ($\delta \mathcal{R} \neq 0$) within the configuration. It is important to stress we are considering local perturbations of density, that can be properly described by any function of compact support, $\delta \rho = \delta \rho(r)$, defined in a closed interval $\Delta r \ll r_b$, where $r_b$ is the total radius of the configuration. 

Accordingly, local density perturbations, $\rho  \rightarrow \rho + \delta \rho$, generate fluctuations in mass, radial pressure, tangential pressure and radial pressure gradient, that can be represented up to linear terms in density fluctuation as: 
\begin{eqnarray} 
 \delta P &=& \frac{\mathrm{d}P}{\mathrm{d} \rho} \delta \rho =  v^2 \delta \rho \,, 
 \label{deltas1} \\   \nonumber \\
 \delta P_\perp &=& \frac{\mathrm{d}P_\perp}{\mathrm{d} \rho} \delta \rho =  v_\perp^2 \delta \rho \,, 
 \label{deltas2} \\   \nonumber \\ 
\delta P' &=& \frac{\mathrm{d} P'}{\mathrm{d} \rho} \delta \rho  =   \frac{\mathrm{d \ }}{\mathrm{d} \rho} \left[ \frac{\mathrm{d} P}{\mathrm{d}r} \right] \delta \rho =   \frac{\mathrm{d}}{\mathrm{d}\rho} \left[ \frac{\mathrm{d}P}{\mathrm{d}\rho} \frac{\mathrm{d}\rho}{\mathrm{d} r} \right]\delta \rho =
 \frac{\mathrm{d}}{\mathrm{d}\rho} \left[v^2 \rho' \right]\delta \rho =
 \frac{1}{ \rho'} \frac{\mathrm{d}}{\mathrm{d}r} \left[v^2 \rho' \right]\delta \rho 
  \label{deltas3} \nonumber 
 \\   \nonumber \\
&=&   \left[  (v^2)' + v^2\frac{\rho''}{\rho'} \right] \delta \rho  \,,  \\  \nonumber \\
\delta m &=& \frac{\mathrm{d} m}{\mathrm{d} \rho} \delta \rho =  \frac{\mathrm{d} m}{\mathrm{d} r} \left( \frac{\mathrm{d} r}{\mathrm{d} \rho} \right) \delta \rho   =  \frac{m'}{ \rho'} \delta \rho = \frac{4 \pi r^2 \rho}{\rho'}\delta \rho   \,.
\label{deltas4} 
\end{eqnarray}
where 
\begin{equation}
 v^{2} = \frac{\mathrm{d} P }{ \mathrm{d} \rho}  \,\, \mathrm{and} \,\, 
 v_\perp^2 = \frac{\mathrm{d} P_\perp }{ \mathrm{d} \rho}\,,
\end{equation}
are the radial and tangential sound speeds, respectively. 

Note that the present perturbation scenario contrasts the original one presented by Herrera and collaborators 
\cite{Herrera1992,DiPriscoEtal1994,DiPriscoHerreraVarela1997}, where fluctuations in density and anisotropy were considered  independent and simultaneous; and it is also different from a previous work \cite{AbreuHernandezNunez2007b} because there pressure gradient were not affected by the density perturbation. 

Following \cite{GonzalezNavarroNunez2017}, we formally expand the quantity $ \mathcal{R}$ emerging from the TOV equation as:
\begin{equation}
 \label{Ranitov}
 \mathcal{R} \equiv \frac{\mathrm{d} P}{\mathrm{d} r} +(\rho +P)\frac{m + 4 \pi r^{3}P}{r(r-2m)} -\frac{2}{r}\left(P_\perp-P \right) \, ,
\end{equation}
as
\begin{equation}
\label{RAniExpanded}
\mathcal{R} \approx \mathcal{R}_{0}(\rho, P, P_\perp, m, P^{\prime}) +
\frac{\partial \mathcal{R}}{ \partial \rho} \delta \rho
+\frac{\partial \mathcal{R}}{ \partial P} \delta P
+ \frac{\partial \mathcal{R}}{\partial P_\perp} \delta P_\perp
+\frac{\partial \mathcal{R}}{ \partial m} \delta m 
+\frac{\partial \mathcal{R}}{ \partial P^{\prime}} \delta P'  \,,
\end{equation}
where $\mathcal{R}_{0}(\rho, P, P_\perp, m, P^{\prime})=0$, because initially the configuration is in equilibrium.  Next, by using (\ref{deltas1})-(\ref{deltas4}) the above equation (\ref{RAniExpanded}) can be reshaped as:
\begin{equation}
 \delta \mathcal{R} \equiv \delta \underbrace{P'}_{\mathcal{R}_p}  + \delta \underbrace{\left[ (\rho +P)\frac{m + 4 \pi r^{3}P}{r(r-2m)} \right]}_{\mathcal{R}_g}  + \delta \underbrace{\left[ 2\frac{P}{r}- 2\frac{P_\perp}{r}\right]}_{\mathcal{R}_a} = \delta \mathcal{R}_p + \delta \mathcal{R}_g +\delta \mathcal{R}_a \, ,
\end{equation}
where it is clear that density perturbations $\delta \rho(r)$ are influencing: the distribution of reacting pressure forces $\mathcal{R}_p$, gravity forces $\mathcal{R}_g$ and anisotropy forces $\mathcal{R}_a$.  Depending on this effect, each perturbed distribution force can contribute in a different way to the change of sign of $\delta \mathcal{R}$: each term can be written as 
\begin{equation}
\label{Fp_p}
\delta \mathcal{R}_p = \left[ \frac{P''}{\rho'} \right] \delta \rho = \left[(v^2)' + v^2\frac{\rho''}{\rho'}\right] \delta \rho
\end{equation}
\begin{equation}
\label{Fp_g}
\delta \mathcal{R}_g = \left[ \frac{ \partial \mathcal{R}_g }{ \partial \rho } + \frac{ \partial \mathcal{R}_g }{ \partial P } v^2 + \frac{ \partial \mathcal{R}_g }{ \partial m } \frac{4 \pi r^2 \rho}{\rho'}\right] \delta \rho \quad \textrm{and}
\end{equation}
\begin{equation}
\label{Fa}
\delta \mathcal{R}_a = 2\left[ \frac{ v^2 - v^2_\perp}{ r } \right] \delta \rho ,
\end{equation}
with
\begin{equation}
\label{Fg_rho}
\frac{\partial \mathcal{R}_g}{\partial \rho}  =  \frac{m+  4 \pi r^3 P }{ r(r - 2m)}\,, \,\,
\frac{\partial \mathcal{R}_g}{\partial P} = \frac{ m + 4 \pi r^3 ( \rho + 2 P )}{r(r-2m)} \quad \textrm{and} \quad
\frac{\partial \mathcal{R}_g}{\partial m}  =  \frac{ (\rho + P)( 1 + 8\pi r^2 P)  }{(r- 2 m )^2 }.
\end{equation}

Notice that if, as in \cite{AbreuHernandezNunez2007b}, the perturbation $\delta \rho$ is constant and does not affect the pressure gradient, we have: $\delta \mathcal{R}_p = 0$, 
\begin{equation}
\label{AbreuEtAl}
\delta \tilde{\mathcal{R}}_g = 2\left[\frac{m+4 \pi r^3 P }{ r(r - 2m)} +\frac{2 \pi r^2}{3} \frac{ (\rho + P)( 1 + 8\pi r^2 P)  }{(  2 m  - r )^2}\right]\delta \rho 
 \,.
\end{equation}
Thus, only anisotropic matter distribution can present cracking instabilities because $\delta \tilde{\mathcal{R}}_g >0$ for all $r$ and the possible change of sign for $ \delta \mathcal{R}$ should emerge from $\delta \mathcal{R}_a$ and the criterion against cracking is written as: 
\begin{equation}
\label{CrackingCriterion}
-1 \leq v^2_\perp -v^2 \leq 0 \quad \Leftrightarrow \quad 
0 \geq \frac{\mathrm{d} P_\perp}{\mathrm{d} r} \geq \frac{\mathrm{d} P}{\mathrm{d} r} \, ;
\end{equation}
more recently this equivalence between the restriction on pressures and velocities was demonstrated in \cite{Ivanov2017} and included as part of the acceptability conditions that have to be considered when building physically reasonable compact object models. We shall discuss these constraints in the next section.

\section{Physical acceptability conditions}
\label{AceptabilityConditions} 
More of the spherically symmetric perfect fluid ``exact solutions'' of Einstein field equations found in the literature are of little physical interest because, in addition to solving the structure equations (\ref{TOVStructure1}) and (\ref{MassStructure2}) for a particular set of equations of state --$P=P(\rho)$ and $P_{\perp}=P_{\perp}(\rho)$-- the physical and metric variables have to comply with several acceptability conditions which, over the years were recently compiled in \cite{Ivanov2017} as:
\begin{enumerate}
\item[C1:] Metric potentials, positive, finite and free from singularities;
\item[C2:] Matching conditions at the surface of the star;
\item[C3:] Decrease of interior redshift $Z$ with the increase of $r$;
\item[C4:] Positive density and pressures;
\item[C5:] Density and pressures having a maximum at the center and decreasing monotonically outwards, with $P_{\perp} \geq P$;
\item[C6:] Energy conditions. Strong (SEC) $\rho \geq P + 2P_{\perp}$ or dominant (DEC) $\rho \geq P$ and $\rho \geq P_{\perp}$;
\item[C7:] Causality Conditions. $0 \leq v^2,v_{\perp}^2 \leq 1$;
\item[C8:] The adiabatic index $\Gamma$ stability criterion as stated in equation (\ref{GammaStability}), which is a consequence of the dynamical stability criterion;
\item[C9:] Stability against cracking as expressed by equation (\ref{CrackingCriterion}); 
\item[C10:] Harrison-Zeldovich-Novikov stability condition:  $\mathrm{d}M(\rho_c)/\mathrm{d}\rho_c > 0$.
\end{enumerate}
Additionally B.V Ivanov \cite{Ivanov2017}, demonstrated that these conditions are not independent and can be condensed in five main inequalities:
\begin{enumerate}
\item $(m/r)' > 0$ which fulfills C1, C2 and C3 conditions;
\item $0 \geq P_{\perp}' \geq P'$ accomplishing: C4, C5, C6 (SEC), C7, C9;
\item $\mu = 2M/r_b \leq 4/5$ executing C6 (DEC);
\item $v^2 \geq 1/3$, implementing C8;
\item and the C10 condition: $\mathrm{d}M(\rho_c)/\mathrm{d}\rho_c > 0$. 
\end{enumerate}

Clearly if we want the configuration to be stable against convection, we should add a sixth condition, i.e. the adiabatic convection stability condition $\rho'' \leq 0$, to the above mentioned set. In the next section, we shall explore its influence on the stability of isotropic and anisotropic models.

\section{Isotropic and anisotropic models}
\label{Models}
In this section we select seven exact solutions --describing isotropic or anisotropic fluid spheres-- which comply with the Ivanov criteria. With this selection we study the effect of convective instability and the reaction of the pressure gradient to density perturbations.  
 
Four of seven density profiles are among most physically reasonable isotropic solutions (Tolman VII \cite{Tolman1939}, Buchdahl-1 \cite{Buchdahl1959}, Mehra \cite{Mehra1966} and Kuchowicz \cite{Kuchowicz1968}) reported in \cite{DelgatyLake1998}. The fifth selection corresponds to a one-parameter family of a generalized Tolman IV solution obtained in \cite{Lake2003} and allows us to exemplify the correlation of convection stability with a decreasing profile of $(v^2)'$. Finally, we study two anisotropic solutions (Gokhroo \& Mehra \cite{GokhrooMehra1994} and Sah \& Chandra  \cite{SahChandra2016}) to illustrate the buoyancy effects in anisotropic matter configurations.   

\subsection{Isotropic solutions}
In addition complying with the Ivanov criteria, isotropic model selection has significant physical interest in describing the interior of compact objects. The isotropic solutions shown in table \ref{TableIsotropicSolution} are: Tolman VII and Mehra are the most frequent parabolic density profiles considered in models of stable neutron stars (see \cite{RaghoonundunHobill2015,BharMuradPant2015,AzamMardanRehman2015,Raghoonundun2016} and reference therein) while Buchdahl's solution \cite{Buchdahl1959,Buchdahl1981} sets limits to the compactness of relativistic spheres. In \cite{DelgatyLake1998} it is reported that the speed of sound for this solution does not decrease  monotonically. We have shown that it can be attained for some particular values of the compactness $\mu = 2M/r_b$. Finally, models of charged spheres are frequently based of Kuchowicz solution \cite{Ivanov2002} and in Mehra's solution \cite{Mehra1966} the density and the speed of sound vanish on the surface. 
\begin{center}
\begin{table}[ht]
\begin{tabular}{|l|c|c|c|}
\toprule[1mm]
& &  &  \\
{\bf Solution} & {\bf Density} $\ \rho$ & {\bf Central density }$ \ \rho_c$ &  $[\rho'']_c$ \\
& &  &  \\ 
\hline \hline
& &  &  \\ 
Tolman VII \cite{Tolman1939} & 
$\dfrac{1}{8\pi A^2}\left[\frac {3A^2}{R^2}-\frac{20\,{r}^{2}}{{A}^{2} }\right]$ & 
${\dfrac{3}{8\pi {R}^{2}}}$ & 
$-{\dfrac {5}{{A}^{4}\pi }}$  \\
& &  &  \\
Buch1 \cite{Buchdahl1959} & 
$\dfrac{3C}{16 \pi}{\dfrac { 3+ C{r}^{2}}{ \left(1+ C{r}^{2} \right) ^{2}}}$ & 
${\dfrac {9\,C}{16\,\pi }}$ &
$-{\dfrac {15\,{C}^{2}}{8\,\pi }}$  \\
&  &  &  \\
Mehra \cite{Mehra1966} & 
$\dfrac{15\mu}{16\pi r_b^2}\left[1 -\left(\frac{r}{r_b }\right)^2 \right]$ &
$\dfrac {15\ \mu}{16\,\pi \,r_b^2}$ & 
$-\dfrac{15\mu}{16\pi r_b^4}$ \\
& &  &  \\ 
Kuch2 III \cite{Kuchowicz1968} & 
$\dfrac {\left[x-x^2 +6 \right]
\left[A F(r) -2  C\right]\mathrm{e}^{-\frac{x}{2} }+
2Ax} { 16\pi (2+ x)  }$ & 
$\dfrac{3\left[ A F(1) -2 C \right]}{16 \pi } $ &
$\dfrac{5A\left[A[1-F(1)]+2\,C \right]}{16\,\pi }$  \\
& &  &  \\ 
\hline \hline
\end{tabular}
\caption{Parabolic density profiles, represented by Tolman VII and Mehra, are used extensively to model stable neutron stars. In these profiles: $A$, $R$ and $C$ are constants to be determined by the boundary conditions. Rational profiles like Buchdahl sets limits to the compactness, $\mu = 2M/r_b$ of relativistic spheres, and $r_b$ is the boundary radius of the matter configuration. Kuchowicz solution models of charged spheres, in this case $x=Ar^2$, the function $F(r) =  \mathrm{Ei}\left(1 +\frac{Ar^2}{2} \right)/\mathrm{e}$ where  $F(0)= \mathrm{Ei}(1)/\mathrm{e}$ and $\mathrm{Ei}$ is the exponential integral function, again, $A$ and $C$ are constants to be determined by the boundary conditions. 
}
\label{TableIsotropicSolution}
\end{table}
\end{center}

{\bf N-parametric isotropic Lake-Tolman IV family of solutions}. 
In reference \cite{Lake2003}, K. Lake  proposed an algorithm based on the choice of a single monotone seed-function, $\mathcal{F}$, to generate all regular static spherically symmetric perfect fluid solutions of  Einstein's equations. Within this scheme we are going to consider the following family of models generated by
\begin{equation}
\label{nulake}
\mathcal{F}(r)=1+Cr^2 \,\, \Rightarrow \, \nu=\frac{N}{2}\ln\left[1+Cr^2\right] \,,
\end{equation}
where $C$ is an arbitrary constant and $N$ is a positive integer that produces an infinite family of analytic solutions. As we can see from the equation (\ref{nulake}), different values of $N$ recover well-known solutions: $N = 1$ corresponds to Tolman IV  solution \cite{Tolman1939}; $N = 3$ represents Heint IIa \cite{Heintzmann1969} solution;  $N = 4$ and $N=5$ are Durg IV and Durg V solutions, respectively \cite{Durgapal1982}. The case $N = 2$ is considered in \cite{LattimerPrakash2005} studying the relationship between the central barionic density and the total mass of observed neutron stars.

As can be easily guessed, the main difficulty of the method lies in how to calculate the two integrals that appearing in equation (4) of \cite{Lake2003}, but fortunately  it is possible in the present case, as can seen in the appendix. 

Thus, with the help of the following auxiliary function
\begin{equation}
\label{Gaux}
\mathcal{G}(r)=1+C{r}^{2}(N+1)\,, 
\end{equation}
we obtain the density profile for any $N$, as
\begin{equation}
\rho(r)= \frac{1}{8\pi r^{2}}\left[ \left(1-\frac{r^2}{2}\,\frac{\Pi_1}{ \mathcal{F}^{N-2}}
\right) \left(2r\Pi_2-1 \right)+1 \right] \,,
\end{equation}
where: 
\begin{eqnarray*}
\Pi_1&=&{\frac {C\left( N-2 \right){N}^{N-2} }{ \left( N+1 \right) ^{N-3}}
\Phi}+\frac{4K}{\mathcal{G}^{\frac{2}{N+1}} }\,,\\
\Pi_2&=&\frac{r}{2\mathcal{F}^{N-2}}
\left[\frac{\left[1- \frac{(\mathcal{F} -1) ( N-2 )}{\mathcal{F}}\right] \Pi_1
 +2(\mathcal{F} -1) \left[{\frac {C ( N-2)  ( N-3 ) }{ ( N+1 )^{N-4} \left(N+ 3 \right) N^2}\Phi}-\frac {4K}{ \mathcal{G}^{\frac{N+3}{N+1}} } \right] }
{1-\frac{(\mathcal{F} -1)}{2C}\,\frac{\Pi_1}{ \mathcal{F}^{N-2}} }\right]
\end{eqnarray*}
and 
\[
\Phi= \mbox{$_2$F$_1$}\left(3-N,\frac{2}{ N+1};\,{\frac {N+3}{N+1}};\,
-{\frac {\mathcal{G} }{N}}\right)\,.
\]
Here, $\mbox{$_2$F$_1$}(a,b;c;d)$ is the hypergeometric function, that for certain special arguments: $(a, b; c; d)$ automatically evaluates to exact values with $K$ a constant.

\begin{table}[ht]
\begin{tabular}{|l|c|c|c|}
\toprule[1mm]
& &  &  \\
{\bf Solution} & {\bf Density} $\ \rho$ & {\bf Central density }$ \ \rho_c$ &  $[\rho'']_c$ \\
& &  &  \\ 
\hline \hline
& &  &  \\ 
FSGM \cite{GokhrooMehra1994} & 
$\dfrac{3\alpha}{8 \pi} \left( 1-{\dfrac {K{r}^{2}}{r_b^2 }} \right) $&
$ {\dfrac {15\,\mu}{8\pi \,r_b^2 \, \left(5-3\,K \right) }}$ & 
$-\dfrac {2 K \rho_c }{ r_b^2 }$ \\
& &  &  \\
Sah \&  Chandra \cite{SahChandra2016} & 
$\dfrac{1}{8 \pi} \left[ 10\,a \left(1-a{r}^{2}\right)^{4}+
\dfrac { 1- \left(1 -a{r}^{2} \right)^{5} }{{r}^{2}} \right] $ & 
$\dfrac {15\,a}{8\,\pi }$ & 
$-\dfrac {25a^2}{2\pi}$  \\
& &  &  \\
\hline \hline
\end{tabular}
\caption{Florides-Stewart-Gokhroo-Mehra (FSGM) and Sah-Chandra profiles satisfy Ivanov criteria. In these profiles: $\alpha$, $K$ and $a$ are constants to be determined by the boundary conditions.}
\label{TabAnisotropicSolutions}
\end{table}

\subsection{Anisotropic solutions}
Local anisotropy (unequal stresses: $P\neq P_{\perp}$) in compact objects can be associated to different physical scenarios --such as phase transition, density inhomogeneity and electromagnetic field, just to mention a few of them-- and has been considered extensively since the work of R. Bowers and E. Liang \cite{BowersLiang1974}.  The unknown physics in the tangential equation of state, $P_{\perp} = P_{\perp}(\rho)$ is partially compensated by using heuristic criteria: geometric, simplicity or any other assumption relating radial and tangential pressures (see \cite{HerreraSantos1997,YagiYunez2017} and references therein). 

Our selection of anisotropic solutions shown in the  table \ref{TabAnisotropicSolutions} are: Florides-Gokhroo-Mehra \cite{Florides1974,Stewart1982,GokhrooMehra1994} and the Sah \&  Chandra \cite{SahChandra2016}. 

The Florides-Gokhroo-Mehra profile was due originally to P.S. Florides \cite{Florides1974}, but also corresponds one of the different solutions considered by Stewart \cite{Stewart1982} and, more recently, by M. K. Gokhroo and A. L. Mehra \cite{GokhrooMehra1994}. The Florides-Stewart-Gokhroo-Mehra (FSGM) solution represents densities and pressures which, under particular circumstances \cite{Martinez1996}, give rise to an equation of state similar to the Bethe-B\"{o}rner-Sato newtonian equation of state for nuclear matter \cite{ShapiroTeukolsky1983,Demianski1985,BetheBornerSato1970}.

\section{Modeling performed and discussion of some results}
\label{Modeling}
\begin{figure}[t]
\begin{center}
\includegraphics[width = 0.46 \textwidth]{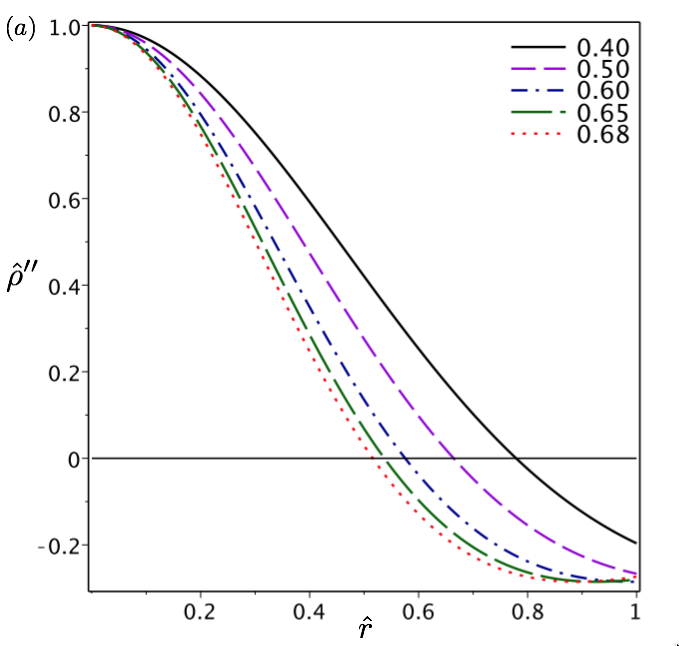} \qquad
\includegraphics[width = 0.46 \textwidth]{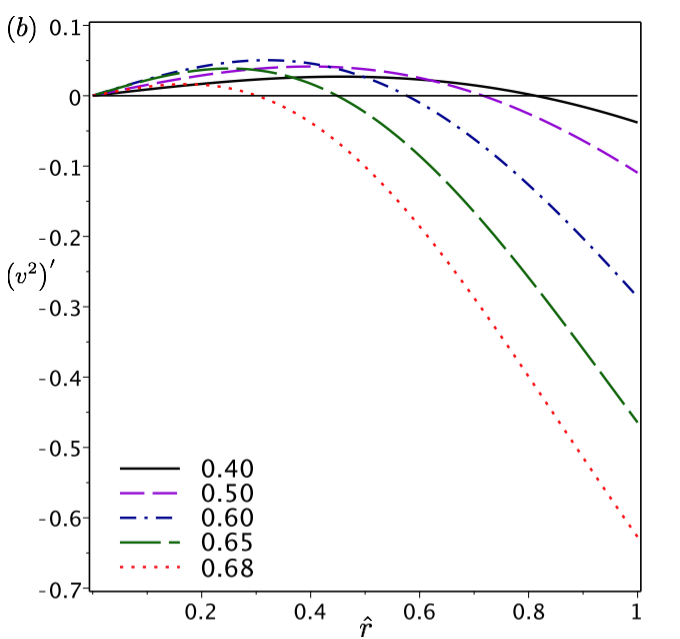}\\
\includegraphics[width = 0.46 \textwidth]{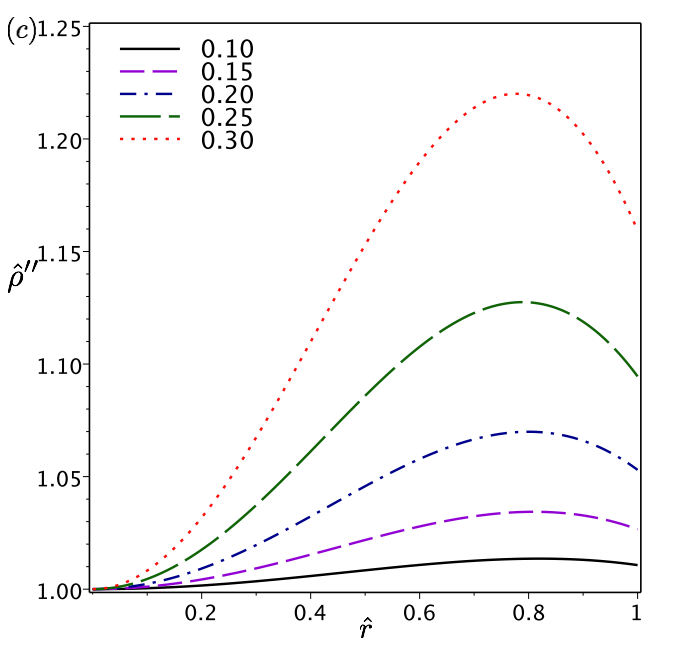} \qquad
\includegraphics[width = 0.46 \textwidth]{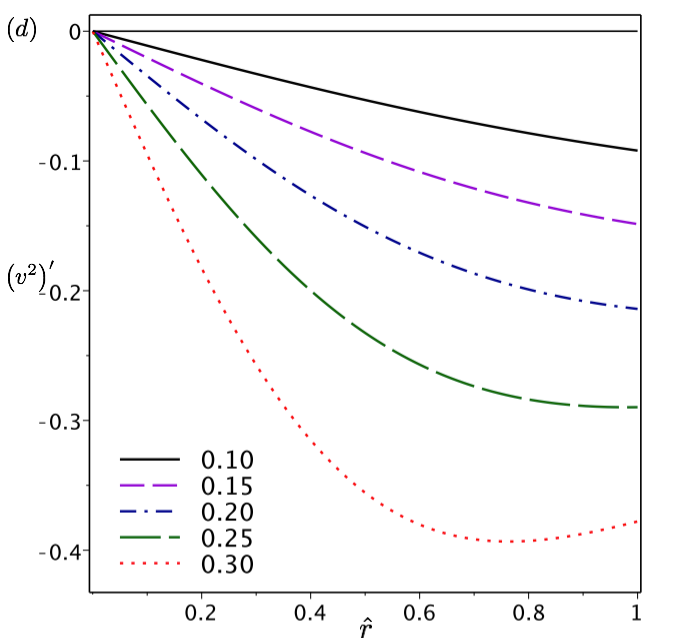}\\ 
\end{center}
\caption{\textbf{Convection for Isotropic Models:} Left plates, (a) and (c), illustrate the normalized buoyancy $\hat {\rho}''~=~\rho''/\rho_{c}''$ vs $\hat{r} = r/r_b$ and right plates, (b) and (d), display the gradient of the sound velocity $(v^2)'$ vs ${\hat r}$ for different values of $\mu=2M/r_b$. Figures (a)-(b) correspond to the Buchdahl model and (c)-(d) to Kuchowicz solution. As it can be appreciated from these plots, Buchdahl model becomes unstable to convective perturbations because $\hat{\rho}''$ changes it sign, while Kuchowicz is stable.}
\label{FigBouyBuchdKuchow}
\end{figure}

\begin{figure}[h]
\begin{center}
\includegraphics[width=0.4\textwidth]{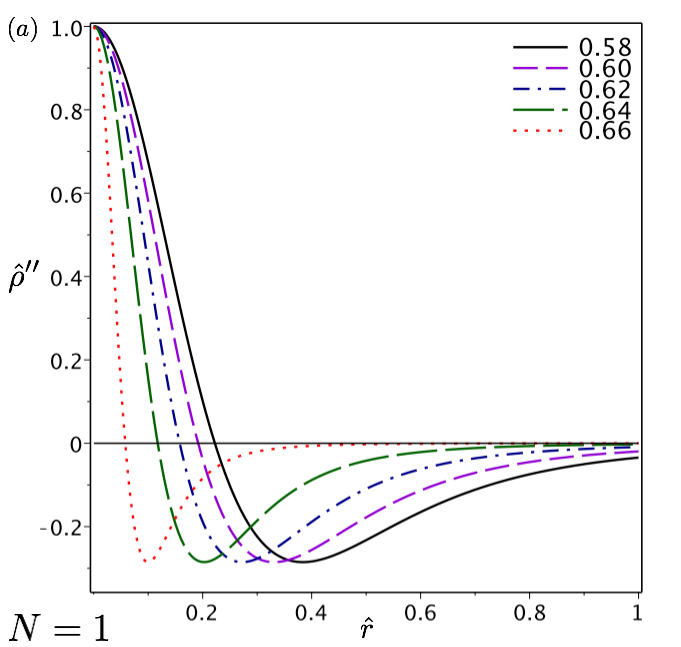}\qquad
\includegraphics[width=0.4\textwidth]{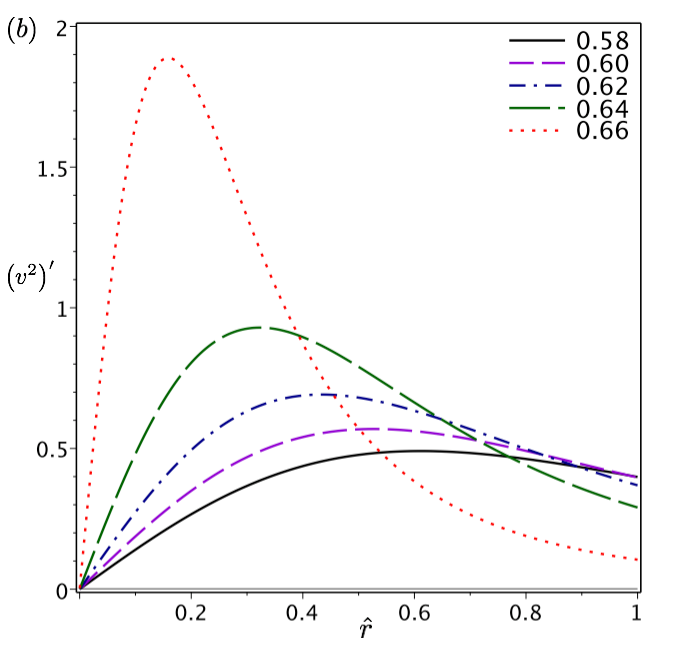}\\
\vspace{0.3cm}
\includegraphics[width=0.4\textwidth]{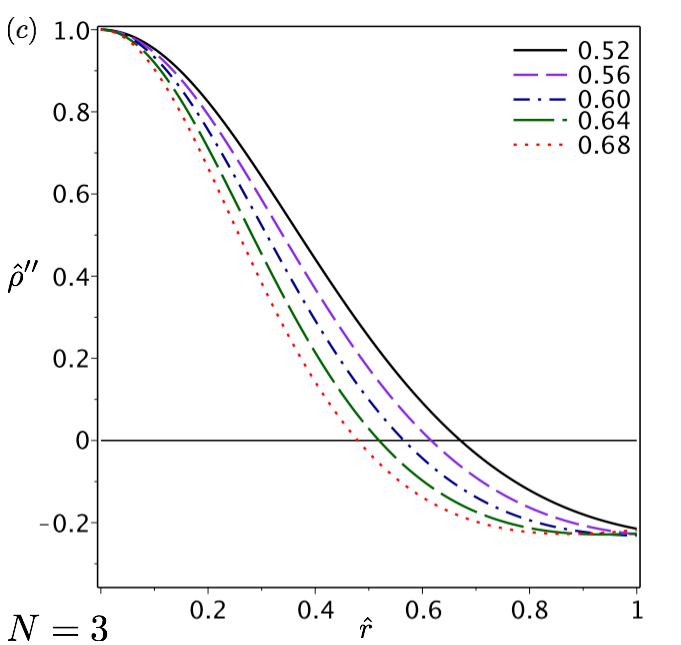} \qquad  
\includegraphics[width=0.4\textwidth]{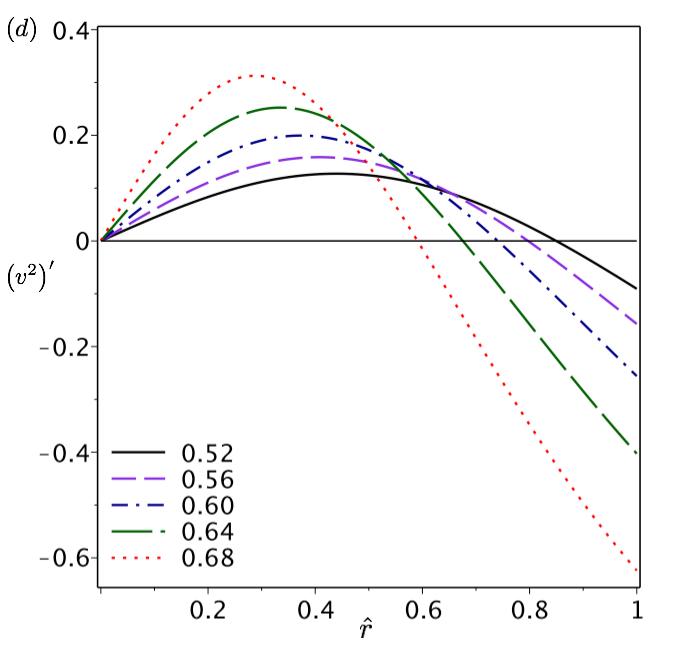}\\
\vspace{0.3cm}
\includegraphics[width=0.4\textwidth]{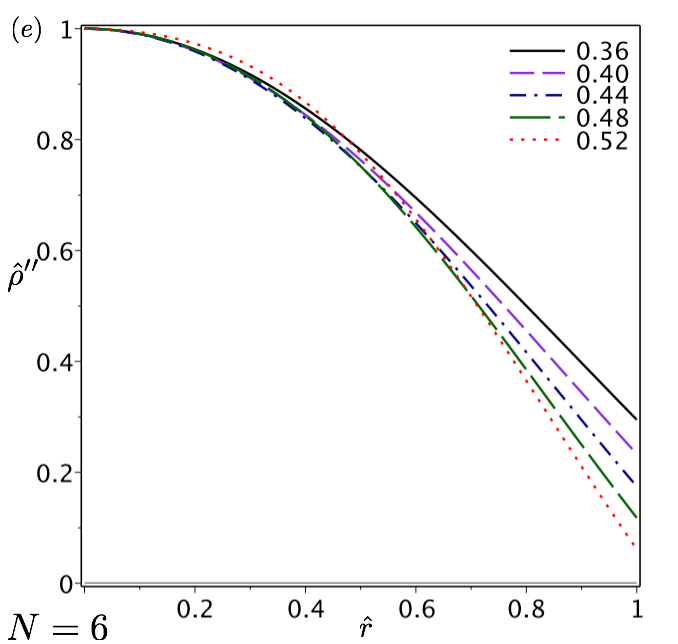} \qquad  
\includegraphics[width=0.4\textwidth]{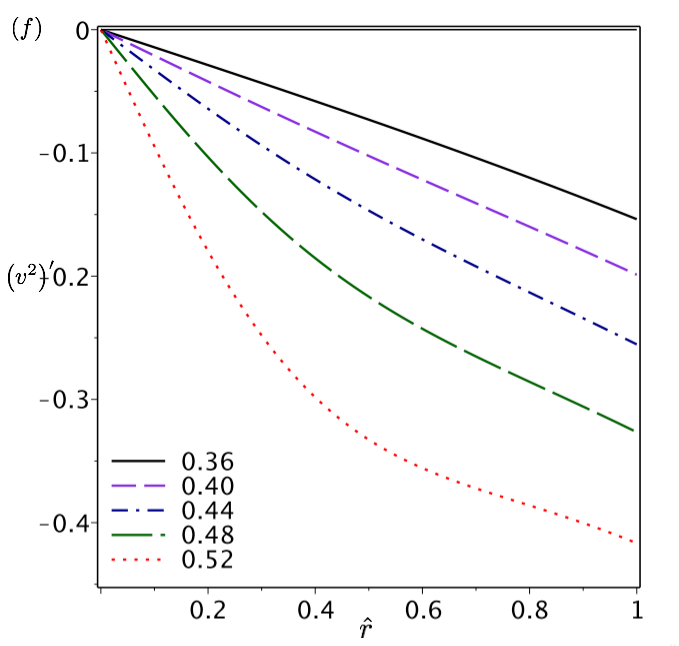} 
\end{center}
\caption{\textbf{Convection for isotropic N-model family:} Normalized buoyancy $\hat{\rho}''$ vs $\hat{r}$  (plates (a)-(c)-(e)) and gradient of the sound velocity $(v^2)'$ vs $\hat{r}$ (plates (b)-(d)-(f)), for different values of $N$ and $\mu= 2M/r_b$. For $N=1$, and $N=3$ (Tolman IV \cite{Tolman1939} and  Heint IIa \cite{Heintzmann1969}, respectively) are unstable and $(v^2)'$ change sign. The $N=6$ Model is stable to convection having $(v^2)'< 0$ for all $\hat{r}$. It is interesting to mention, that for this family of solutions, when $N$ increases, we obtain more convective stable models.}
\label{FigBouyNFamily}
\end{figure}

\begin{figure}[h]
\begin{center}
\includegraphics[width = 0.32 \textwidth]{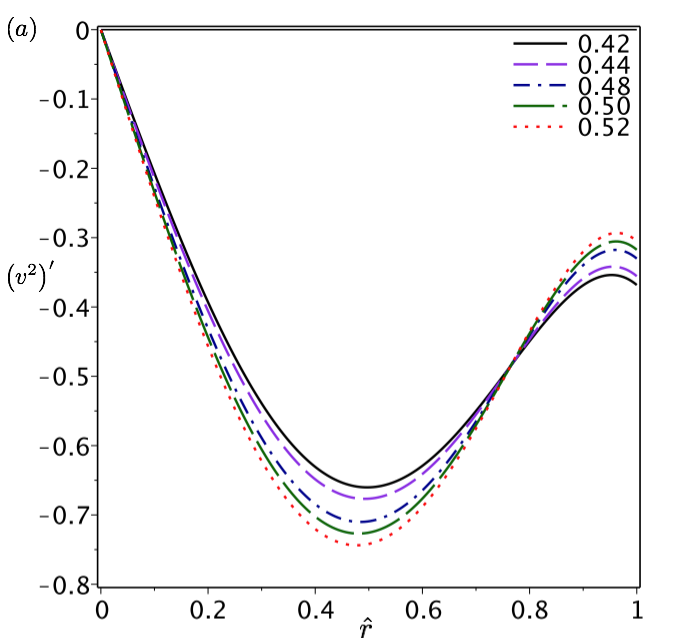}\,\,
\includegraphics[width = 0.32 \textwidth]{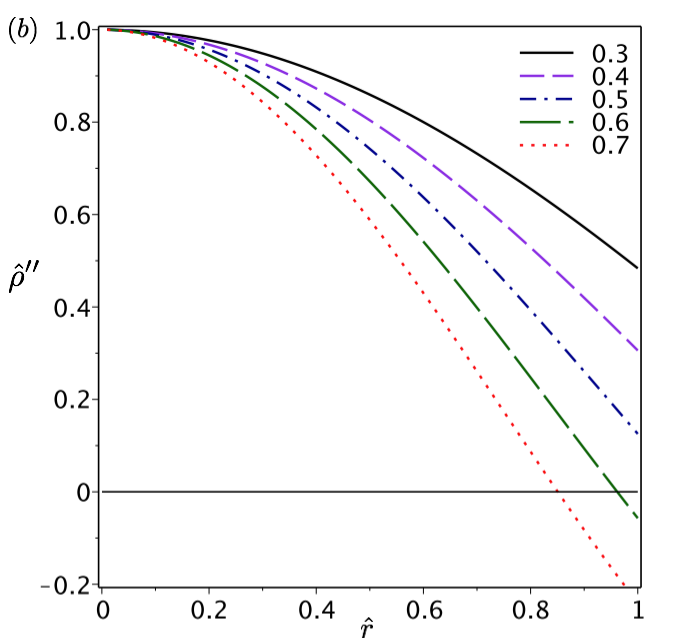} \,\,
\includegraphics[width = 0.32 \textwidth]{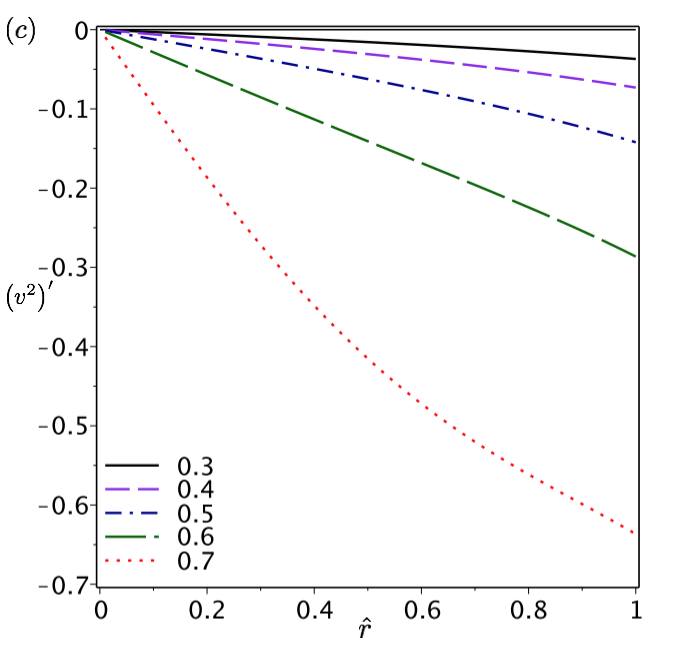}
\end{center}
\caption{\textbf{Convection for anisotropic models:} The
plate (a) illustrates, for different values of $\mu= 2M/r_b$, the gradient of the sound velocity $(v^2)'$ for the FSGM model  while Sah \& Chandra's model, in plates (b) and (c), illustrate the buoyancy ${\rho}^{\prime\prime}$ and $(v^2)'$. Clearly, FSGM has a parabolic density profile and it is stable to convection. Notice in the case of $\mu = 0.7, 0.6$ Sah \& Chandra's solution is unstable in the outer parts of the configuration and $(v^2)'$ is negative. This is because theses models do not comply equation (\ref{Cond1Convection}).
}
\label{FigBouyAniModels}
\end{figure}

The first effect to be analyzed is the convective instability and its relation to the sign of the gradient of the speed of sound, $(v^2)'$, which are illustrated in Figures \ref{FigBouyBuchdKuchow}, \ref{FigBouyNFamily} and \ref{FigBouyAniModels}. We show this effect by plotting the normalized buoyancy $\hat {\rho}'' =\rho''/\rho_{c}''$ vs $\hat{r} = r/r_b$ and the corresponding gradient $(v^2)'$ vs $\hat{r}$. The first two figures display the convective stability for isotropic models while Figure \ref{FigBouyAniModels} illustrates this property for anisotropic spheres.   

As we have mentioned before, parabolic density profiles --represented by Tolman VII, Mehra and Florides-Stewart-Gokhroo-Mehra in Tables \ref{TableIsotropicSolution} and \ref{TabAnisotropicSolutions}-- are stable to convection because they have constant $\rho''< 0$. The stability for the other isotropic models having rational density profiles is presented in Figure \ref{FigBouyBuchdKuchow}, where the Buchdahl model becomes unstable because $\rho''/\rho_c$ changes sign, while Kuchowicz is stable against convective perturbations due to $\rho''/\rho_c > 0$ for all $r$.  

In Figure \ref{FigBouyAniModels} we illustrate the convective instability for anisotropic matter configurations. Again, models that are stable with the Ivanov criteria are revealed unstable for convection. This is the case of then Sah \& Chandra model which is unstable against convective perturbation, but the Florides-Stewart-Gokhroo-Mehra solution is stable because it has a parabolic density profile.

In all the isotropic models analyzed we found an interesting correlation between the stability against convective perturbation and the sign of $(v^2)'$. In particular, in the regions where matter configuration has $\rho'' < 0$ then we have $v^2$ as a monotonous decreasing function. And this is more evident in Figure \ref{FigBouyNFamily} where we have ploted the buoyancy and the gradient ($v^2$)'.  We found unstable configurations for $N=1$, Tolman IV \cite{Tolman1939} and for $N=3$ Heint IIa \cite{Heintzmann1969} with $(v^2)'$ changing sign within the configuration and for $N=6$ a convective stable configuration having $(v^2)'< 0$ for all $\hat{r}$.   
This can be easily understood in those regions where $P''<0$ because
\begin{equation}
\label{Cond1Convection}
\rho'' = -\frac{(v^2)'}{(v^2)^2}P'+ \frac{P''}{v^2}\,,
\,\, \textrm{thus, if} \,\, P'' < 0 \; \wedge \;  (v^2)' < 0 \,\, \Rightarrow \,\, \rho'' < 0\,.
\end{equation}
Therefore, a distribution having a monotonically decreasing concave pressure profile, i.e. $P' < 0$ and $P'' < 0$, will be stable against convection, if the sound velocity  monotonically decreases outward from the configuration.  It is interesting to see that when it happens we have 
\begin{equation}
\label{Cond2Convection}
(v^2)' = \frac{\partial^2 P}{\partial \rho^2}\rho'\,, \,\, \textrm{thus,} \,\, \rho' < 0 \,\,\Rightarrow \,\, (v^2)' < 0 \quad \textrm{if } \; \frac{\partial^2 P}{\partial \rho^2} > 0\, ,
\end{equation}
and this last condition is not only met by the model we have considered here, but also by any relativistic polytropic equation of $P = K \rho^{\gamma}$ or $P = (\gamma -1)\rho$ and by other several more realistic numeric equation of state for ultradense matter (see \cite{OzelFreire2016}, and references therein).

\begin{figure}[h]
\begin{center}
\includegraphics[width=0.4\textwidth]{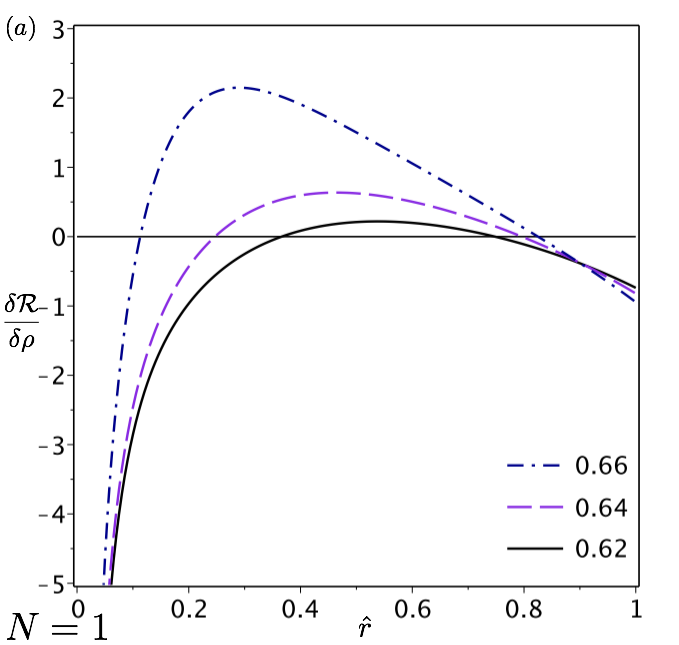}\qquad
\includegraphics[width=0.4\textwidth]{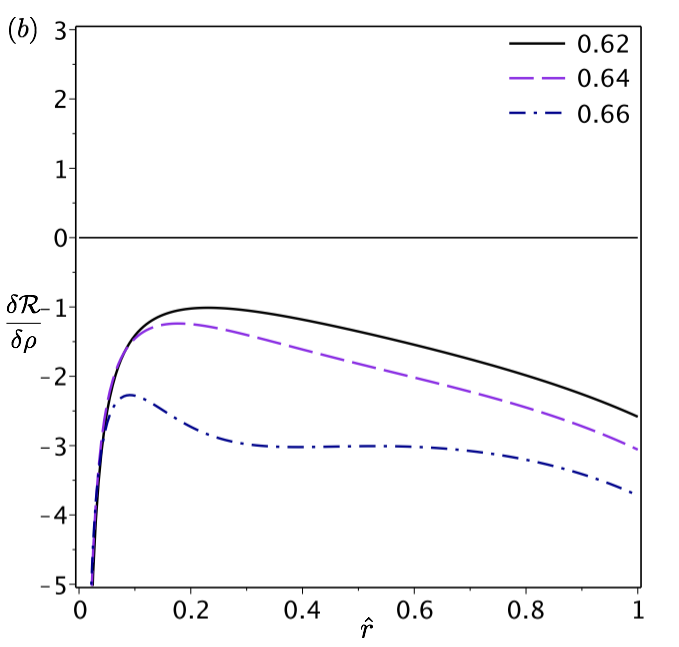}\\ 
\vspace{0.3cm}
\includegraphics[width=0.4\textwidth]{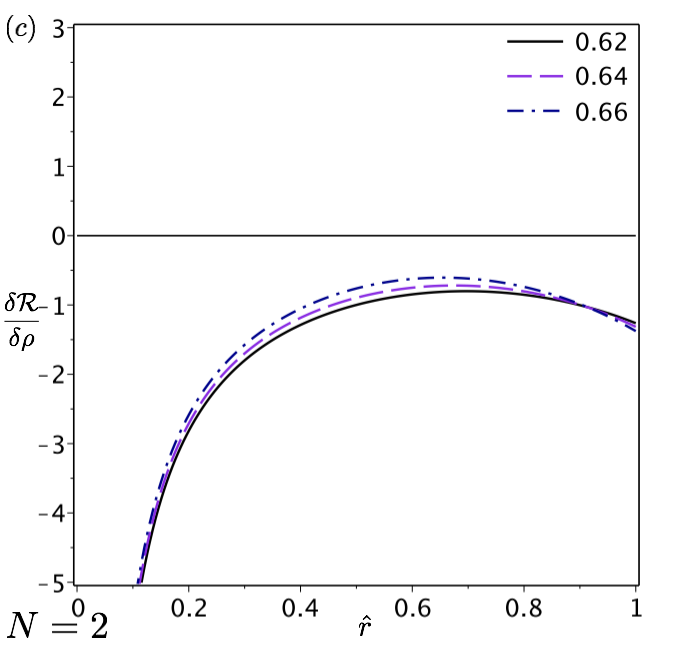} \qquad  
\includegraphics[width=0.4\textwidth]{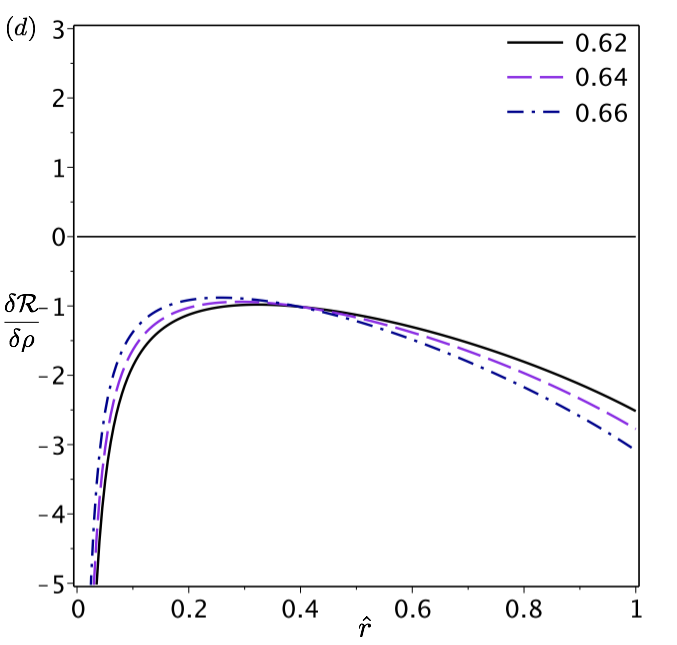}
\end{center}
\caption{\textbf{Stability against cracking, isotropic  $N$-models:} In these figures we show, the stabilizing effect of the reaction of pressure gradient to density perturbation. In these plots we present ${\delta \mathcal{R}}/{ \delta \rho}$ for two members of the Lake $N$-Family. As in the previous case this stabilizing effect is only present for $N=1$, for  $N\geq 2$ no instability appears. Again, the more $N$ increases the most stable the configuration is.
}
\label{FigCrackNFamily}
\end{figure}

\begin{figure}[t]
\begin{center}
\includegraphics[width=0.4 \textwidth]{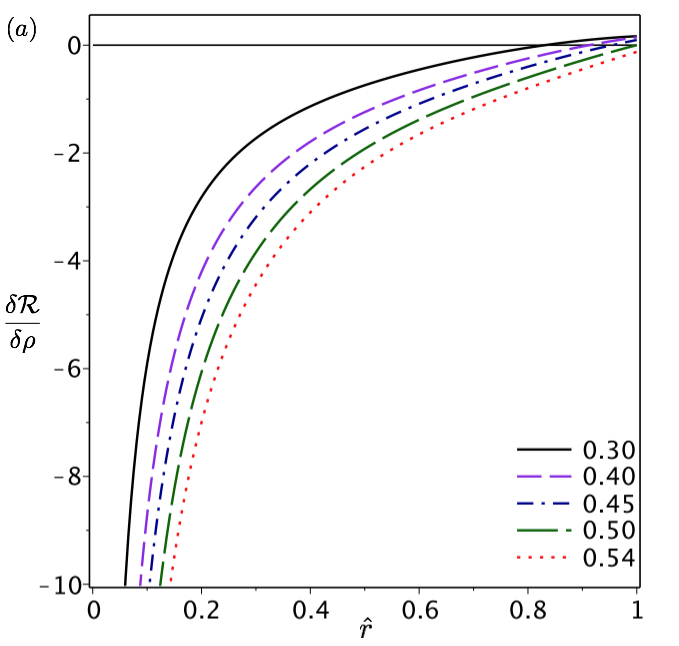}\qquad
\includegraphics[width=0.4 \textwidth]{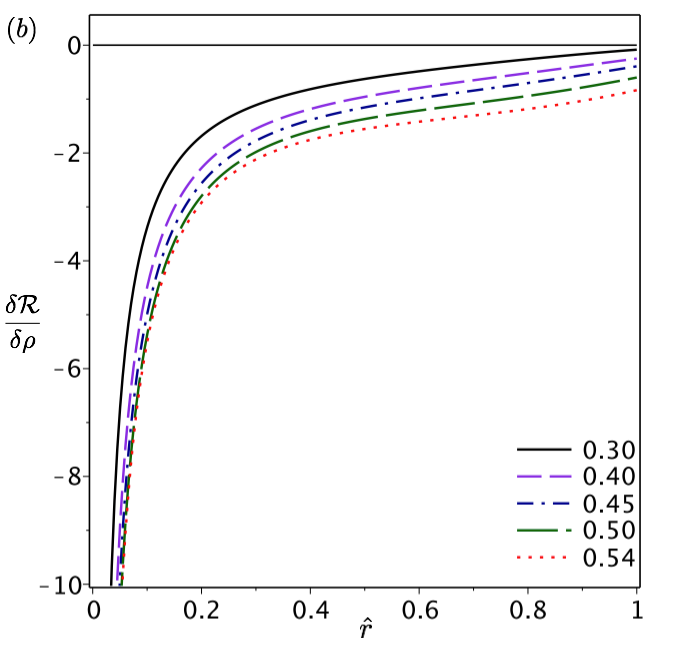}\\ 
\includegraphics[width=0.4 \textwidth]{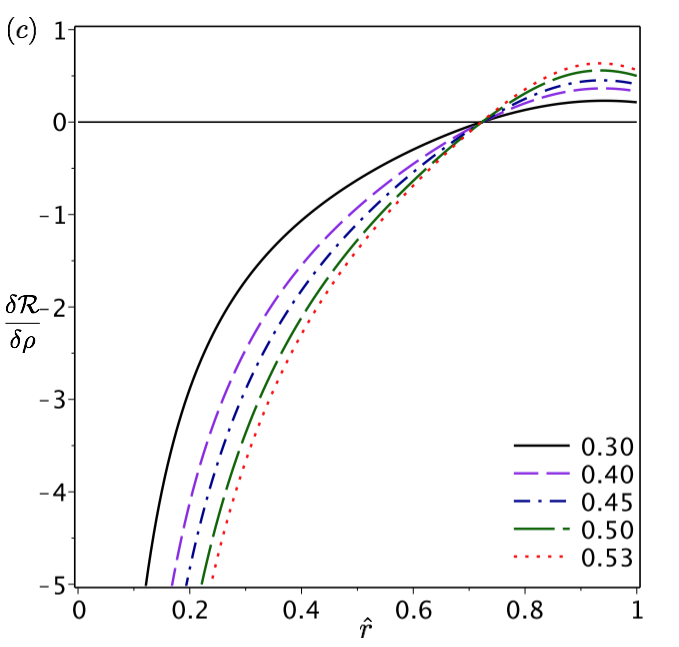}\qquad
\includegraphics[width=0.4 \textwidth]{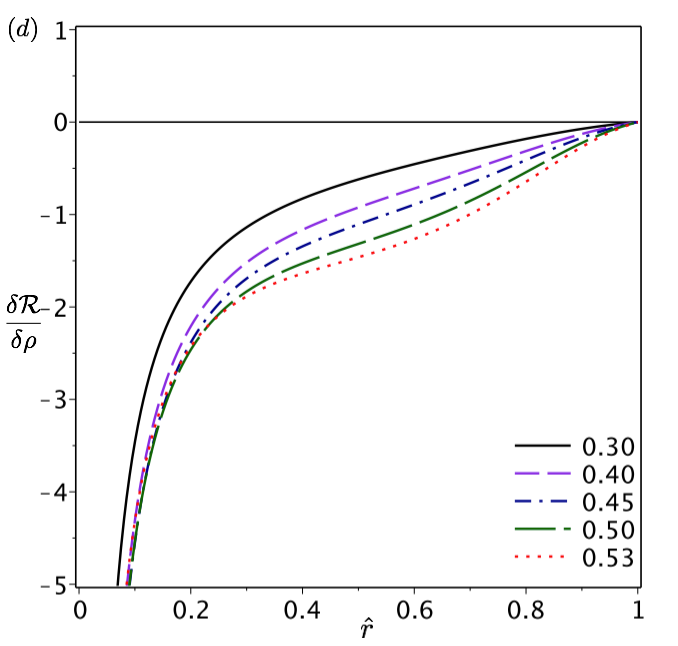}\\ 
\includegraphics[width=0.4 \textwidth]{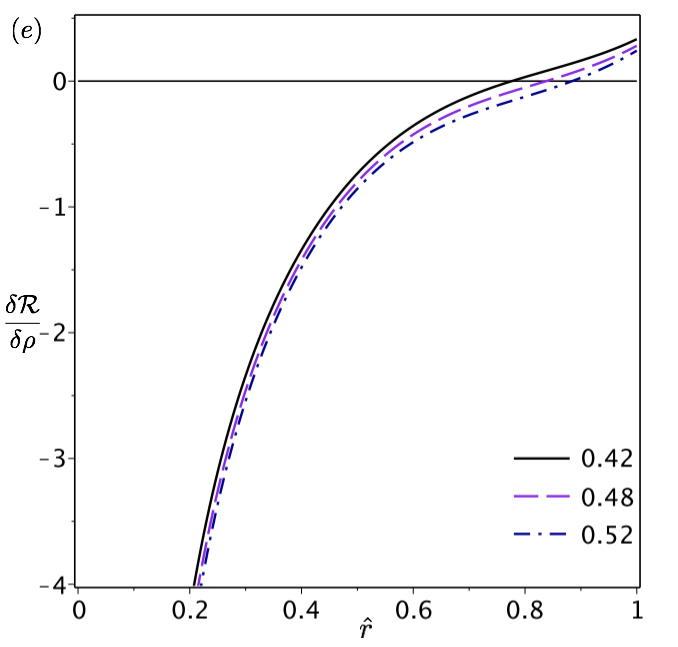}\qquad
\includegraphics[width=0.4 \textwidth]{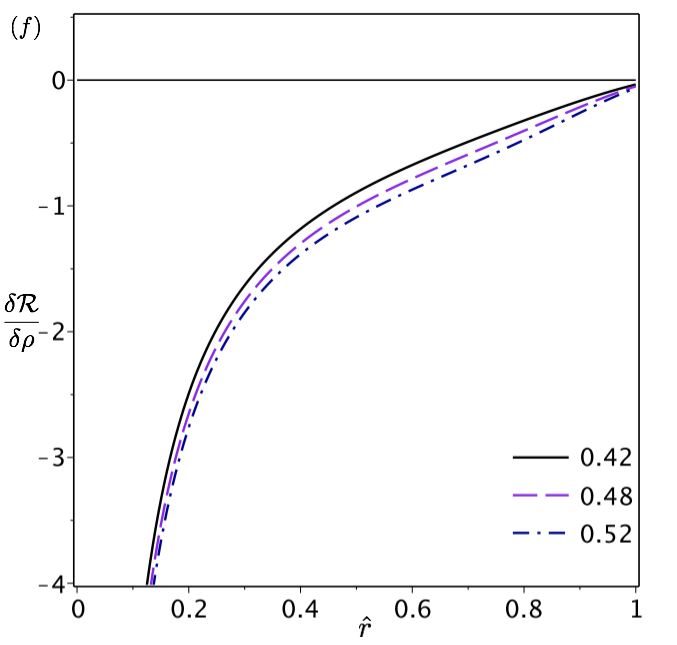}\\ 
\end{center}
\caption{\textbf{Stability to cracking in isotropic \& anisotropic models:} In these plots we show, for different values of $\mu= 2M/r_b$, the stabilizing effect of the reaction of pressure gradient to density perturbation. Plates (a)-(b) correspond to Tolman VII models; plates (c)-(d) to the Mehra models and (e)-(f) to the FSGM model. We compare  $\delta \mathcal{R}/ \delta \rho$ as a function of $\hat{r}$ when the reaction of the pressure gradient to density perturbation are considered (plates (b), (d) and (f)) and when they are not (plates (a), (c) and (e)). It clear that these models are unstable to cracking when the reaction of the force distribution  is not considered, i.e. $\delta \mathcal{R}_p =0$. Notice that, contrary to what was shown in \cite{GonzalezNavarroNunez2017} Mehra models are stable when $\delta \mathcal{R}_p$ is taken into account. This may happen because a possible mismatch in pressure and density for this solution in this reference.
}
\label{FigCrackToVIIMehra}
\end{figure}

The other effect to be discussed is the stability induced by the reaction of the pressure gradient to density perturbations. This is shown in Figures \ref{FigCrackNFamily} and \ref{FigCrackToVIIMehra}, where we compare the perturbation on the total force distribution when the pressure gradient is perturbed and when it is not. In these figures we plot $\delta \mathcal{R}_p \neq 0$, and $\delta \mathcal{R}_p = 0$, respectively. The first case, $\delta \mathcal{R}_p \neq 0$, represents the recent cracking scheme of  Gonzalez-Navarro-Nunez \cite{GonzalezNavarroNunez2015,GonzalezNavarroNunez2017} while the second one corresponds to a variation of the previous work of Abreu-Hernandez-Nunez \cite{AbreuHernandezNunez2007b}. As we have stressed above, in the first case it is assumed a non-constant density perturbation $\delta \rho = \delta \rho(r)$, which leads to the factor $ \frac{4 \pi r^2 \rho}{\rho'} \leq 0 $ in the third term in the perturbation to the gravitational force distribution $\delta \mathcal{R}_g$ in equation (\ref{Fp_g}),  which may cause cracking instabilities.  

As it is evident for the models considered, if the pressure gradient is not perturbed, i.e. $\delta \mathcal{R}_p~=~0$, $\delta \mathcal{R}$ may change its sign and potential cracking instabilities may appear. On the other hand, if the gradient reacts to the perturbation, $\delta \mathcal{R}_p \neq 0$, we find that $\delta \mathcal{R}$ does not change sign and the matter configuration becomes stable against cracking. This tendency to make models cracking-stable if the the pressure gradient reacts to the density, was previously reported incidentally in reference \cite{GonzalezNavarroNunez2017}. 

This induced stability can be understood if we shape the perturbation of the hydrostatic equation. Clearly, for the isotropic case, $\mathcal{R}_a =\delta \mathcal{R}_a = 0$ and when $\delta \mathcal{R}_p = 0$, the cracking instability emerges only from the effect of the perturbation on the gravitational force distribution, $\delta \mathcal{R}_g$ and particularly from the third term, which is always negative. If $\delta \mathcal{R}_p \neq 0$ and $P'' > 0$, the reaction of the pressure gradient to the density perturbation can neutralize the effect of the negative sign of the above mentioned gravitational term.   

\section{Conclusions and final remarks}
\label{FinalRemarks}
Stability is a key concept when considering self-gravitating stellar models, because only those in stable equilibrium are of astrophysical interest. A we have stated above, in addition to solving the structure equations (\ref{TOVStructure1}) and (\ref{MassStructure2}) for a particular set of equations of state: $P=P(\rho)$ and $P_{\perp}=P_{\perp}(\rho)$, the emerging physical variables have to comply with the several acceptability conditions stated in Section \ref{AceptabilityConditions}.  

The stability of a spherical star against convection has almost been forgotten in most of the stability analysis. It is a very simple criterion which implements the Archimedes principle in any hydrostatic matter configuration \cite{Bondi1964B,Thorne1966,Kovetz1967}. This criterion has proven to be interesting because several reasonable models become unstable against convection and should be included in the acceptability criteria to guarantee physically interesting models of compact objects. 

In this work we have shown that:
\begin{enumerate}
\item A density profiles with its second derivative with respect to the radial marker less or equal than zero, $\rho'' < 0$, will be stable against convective motions. This is a very simple criterion to identify potential convection instabilities within spherical matter configurations and it should be added to the above mentioned acceptability Ivanov criteria.  

\item A decreasing concave pressure profile $P(r)$, i.e. $P' < 0$ and $P'' < 0$, will be stable against convection if the radial sound velocity decreases monotonically  outward in spherically matter configurations. Equivalently, if $ \frac{\partial^2 P}{\partial \rho^2} > 0$, then $(v^2)' < 0$ implies stability against convective motions. It should be pointed out that to illustrate this effect we have implemented a new family of exact solution by using the method propose by in reference \cite{Lake2003}.

\item From (\ref{RAniExpanded}) we obtained the possible sources of cracking: the reaction of the pressure gradient $\delta \mathcal{R}_p$; the perturbation to the gravitational force distribution $\delta \mathcal{R}_g$ and the perturbation of the anisotropy $\delta \mathcal{R}_a$.  Convection may cause cracking instability only when the perturbation of the pressure gradient is considered, i.e. $\delta \mathcal{R}_p \neq 0$, in this case cracking density perturbations affect the pressure gradient, depending on the values of $(v^2)'$ and $v^2\frac{\rho''}{\rho'}$.   

\item Isotropic and anisotropic models considered can be unstable to cracking when the reaction of the pressure gradient is neglected, i.e. $\delta \mathcal{R}_p = 0$, but if taken into account, the instabilities may vanish. Thus, there is a stabilizing effect against cracking, when the perturbation gradient is affected by density perturbation,  $\delta \mathcal{R}_p \neq 0$. 
\end{enumerate} 

Local perturbed schemes are based on the reaction of the fluid variables to a density fluctuation that drives the system out of its equilibrium, i.e. we are exclusively considering perturbations under which the system is dynamically unstable. One way to achieve this is to assume for a barotropic fluid,  that pressure gradients are not affected and we have shown that this occurs in several of the models considered. Convection contributes to the pressure gradient reaction but instead of developing or increasing crancking, it stabilizes the configuration.  

Finally, it should be stressed that the results we have presented of possible instabilities for local perturbation schemes (convection and/or cracking) have to be considered as tendencies that could lead potential evolution of fluids within a relativistic matter configuration, but this should emerge from the full integration of the Einstein Equations.

\section*{Acknowledgments}
We gratefully acknowledge the financial support of the Vicerrector\'ia de Investigaci\'on y Extensi\'on de la Universidad Industrial de  Santander and the financial support provided by COLCIENCIAS under Grant No. 8863.

\section*{Appendix}
As we already mentioned, K. Lake \cite{Lake2003} formulated a method to generate infinite static solutions from a seed function for the case of perfect fluid. This method was also extended to the anisotropic case in \cite{HerreraOspinoDiPrisco2008}. For the isotropic case we have that given the function $\nu(r)$ then $m(r)$ can be integrated as it is shown below
\[
f_1=\frac{r\left[r\left(\nu''+(\nu')^2 \right) -\nu' \right]}{r\nu'+1}\,,\,\,
f_2=\frac{1}{r}\left[2f_1-\frac{r \nu'+3}{r\nu'+1}\right] \,, \,\, 
f_3= e^{\int f_2 \mathrm{d}r} 
\,\, \mbox{and}\,\, 
f_4=\int f_1 f_3 \mathrm{d}r\,.
\]
So:
\begin{equation}
m(r)=\frac{f_4  + K}{f_3}
\,\, \Rightarrow \,\,
e^{-2\lambda}=1-\frac{2m}{r} \,.
\label{emedenu}
\end{equation}
where $K$ is a constant.

With the functions (\ref{nulake}) and (\ref{Gaux}) we have
\begin{eqnarray*}
f_1(r)&=&{\frac { N\left( N-2 \right) {C}^{2}{r}^{4}}
{ \mathcal{F} \ \mathcal{G} }}\,,\quad 
f_2(r)=\frac{1}{r}{\frac {{C}^{2}{r}^{4}\left( 2\,N+1 \right)  \left( N-3 \right)-C{r}^{2}(N+6)-3}{ \mathcal{F} \ \mathcal{G}}} \,, \\
f_3(r)&=& \frac{1}{r^3}\frac {  {\mathcal{G}}^{ \frac{2}{ N+1}} }
{ \mathcal{F}^{2-N}}  \nonumber \,, \quad  
f_4(r)=\frac {C N^{N-2}\left( N-2 \right) 
\mathcal{G}^{\frac{2}{ N+1}}}{4\left( N+1 \right)^{N-3}}  \Phi \,,
\end{eqnarray*}
where
\[
\Phi(r)= \mbox{$_2$F$_1$}\left(3-N,\frac{2}{ N+1};\,{\frac {N+3}{N+1}};\,
-{\frac {\mathcal{G} }{N}}\right) \,,
\]
and $\mbox{$_2$F$_1$}(a,b;c;d)$ is the hypergeometric function that for certain special arguments, $\mbox{$_2$F$_1$}(a,b;c;d)$ automatically evaluates to exact values.

The resulting mass function is:
\[
m(r)=\frac{r^3}{4}\frac{1}{\mathcal{F}^{N-2}}\left[{\frac {C\left( N-2 \right){N}^{N-2} }{ \left( N+1 \right) ^{N-3}}
\Phi}+\frac{4K}{\mathcal{G}^{\frac{2}{N+1}} }\right]\,.
\]
If we define
\[
\Pi_1(r)\equiv{\frac {C\left( N-2 \right){N}^{N-2} }{ \left( N+1 \right) ^{N-3}}
\Phi}+\frac{4K}{\mathcal{G}^{\frac{2}{N+1}} }\,,
\]
then the density can be written as
\[
\rho(r)= \frac{1}{8\pi r^{2}}\left[ \left(1-\frac{r^2}{2}\,\frac{\Pi_1}{ \mathcal{F}^{N-2}}
\right) \left(2r\Pi_2-1 \right)+1 \right]\,,
\]
where
\[
\Pi_2(r)=\frac{r}{2\mathcal{F}^{N-2}}
\left[\frac{\left[1- \frac{C{r}^{2} \left( N-2\right)}{\mathcal{F}}\right] \Pi_1
 +2Cr^2 \left[{\frac {C \left( N-2 \right)  \left( N-3 \right) }{ \left( N+1 \right) ^{N-4} \left(N+ 3 \right) N^2}\Phi}-\frac {4K}{ \mathcal{G}^{\frac{N+3}{N+1}} } \right] }
{1-\frac{r^2}{2}\,\frac{\Pi_1}{ \mathcal{F}^{N-2}} }\right]\,.
\]
The resulting pressure is then
\[
P(r)=\rho-\frac{1}{4\pi r^{2}}{\left[1-\frac{r^2}{2}\,\frac{\Pi_1}{ \mathcal{F}^{N-2}}\right]}
\left[r\Pi_2 -{\frac {NCr^2}{\mathcal{F}}}+\frac{1}{1-\frac{r^2}{2}\,\frac{\Pi_1}{ \mathcal{F}^{N-2}}}-1\right]\,.
\]

Because of the boundary condition $P_b\equiv P(r_b)=0$, it follows that the constant $K$ is:
\[
K=\frac{CN}{4}\,
\frac{\left( 1+Cr_b^2 \right)^{N-2}
\left[1+Cr_b^2(N+1) \right]^{\frac{2}{N+1}}} {1+Cr_b^2(2N+1) } 
\left[ 4-
{\frac{N^{N-3} \left( N-2 \right)  \ \Pi_3} 
{\left( N+1 \right)^{N-3}  \left(1+ C r_b^2 \right)^N }} \Phi_b
\right]
\]
Where:
\[
\Pi_3= {C}^{3}r_b^6\left( 2\,N+1 \right)  +{C}^{2}r_b^4 \left( 4\,N+3 \right) +{C}r_b^2\left( 2\,N+3 \right) +1 
\]
and
\[
\Phi_b= \mbox{$_2$F$_1$}\left(3-N,\frac{2}{ N+1};\,{\frac {N+3}{N+1}};\,
-{\frac {\mathcal{G}_b }{N}}\right) \,.
\]

The constant $C$ is obtained from the condition $m(r_b)=M$
\[
C=\frac {M}{ r_b^2\left[N r_b -(2N+1)\,M \right] } \,.
\]

These models are always regular in the center, the central density $\rho_c \equiv\rho(0)$ is
\[
\rho_c=\frac{3}{16\pi}  \left[\frac{C {\left( N-2 \right) N^{N-2}}}{\left( N+1 \right) ^{N-3}}
\ \Phi_2 + 4\,K 
 \right]\,,
\]
and the central pressure
\[
P_c=\rho_c-{\frac {1}{4\pi } \left[ 
\frac {C \left( N-2 \right) N^{N-2}}{(N+1)^{N-3}}   \Phi_2 + 4\,K-{N}C\right]} = -{\frac {1}{4\pi } \left[ \frac {C \left( N-2 \right) N^{N-2}}{4(N+1)^{N-3}}  \Phi_2 + K-{N}C\right]}\,,
\]
where
\[
\Phi_2={\mbox{$_2$F$_1$}\left(3-N,\frac{2}{N+1};\,{\frac {N+3}{N+1}};-\frac{1}{N}\right)}\,.
\]


\bibliography{BiblioLN180430}

\end{document}